\documentclass[twocolumn,showpacs,preprintnumbers,amsmath,amssymb]{revtex4}

\usepackage{graphicx,epsf}

\usepackage{dcolumn}

\usepackage{bm}

\begin{document}

\title{Disorder-induced microscopic magnetic memory}

\author{M.S. Pierce$^{1}$, C.R. Buechler$^{1}$, L.B. Sorensen$^{1}$, 
J.J. Turner$^{2}$, S.D. Kevan$^{2}$,
E.A. Jagla$^{3}$,
J.M. Deutsch$^{4}$,T. Mai$^{4}$, O. Narayan$^{4}$,
J.E. Davies$^{5}$, K. Liu$^{5}$,
J. Hunter Dunn$^{6}$,
K.M. Chesnel$^{7}$, J.B. Kortright$^{7}$,
O. Hellwig$^{8\dag}$, and E. E. Fullerton$^{8}$.
}

\address{$^{1}$Department of Physics, University of
Washington, Seattle, Washington 98195, USA}
\address{$^{2}$Department of Physics, University of 
Oregon, Eugene, Oregon 97403, USA}
\address{$^{3}$The Abdus Salam International Centre
for Theoretical Physics, Trieste, Italy}
\address{$^{4}$Department of Physics, University of
California, Santa Cruz, California 95064, USA}
\address{$^{5}$Department of Physics, University of
California, Davis, California 95616, USA}
\address{$^{6}$MAX Laboratory,
Box 118, SE-221 00 Lund, Sweden}
\address{$^{7}$Lawrence Berkeley National Laboratory,
Berkeley, California 94720, USA}
\address{$^{8}$Hitachi Global Storage
Technologies, San Jose, California 95120, USA}

\date{\today}

\begin{abstract}

Using coherent x-ray speckle metrology,
we have measured the influence of disorder
on major loop return point memory (RPM) and
complementary point memory (CPM) for a series of
perpendicular anisotropy Co/Pt multilayer films.
In the low disorder limit,
the domain structures show no memory with field
cycling---no RPM and no CPM.
With increasing disorder, we observe the onset and the saturation of
both the RPM and the CPM.
These results provide the first direct
ensemble-sensitive experimental
study of the effects of varying disorder on microscopic magnetic memory 
and are compared against the predictions of existing theories.

\end{abstract}


\maketitle

\vskip2pc

\narrowtext

The magnetic recording industry
deliberately introduces carefully controlled disorder into its
materials to obtain the desired hysteretic behavior.
Over the past forty years, such magnetic hardening has developed into a high art form \cite{1}.
Although we do not yet have a
satisfactory
{\it{fundamental
microscopic}}
description of magnetic
hysteresis,
beautiful theories of
magnetic memory
based on random
microscopic disorder have been
developed
over the past ten years \cite{2}.
How well do these theories
compare with
experimental results?
To address this question, we have deliberately introduced increasing degrees of
disorder into a series of thin multilayer perpendicular magnetic
materials and studied how the 
configuration of the magnetic domains evolve as
these systems are cycled around their major hysteresis loops.

Until very recently \cite{3}, our best information about the repeatability of the 
domain configurations was deduced from the associated
magnetic avalanches via the Barkhausen noise \cite{4}.
Our best current microscopic theories \cite{2}
were built on the microscopic information provided
by Barkhausen measurements and on the macroscopic information
provided by classical magnetometry measurements.
If the Barkhausen noise repeats perfectly for every cycle of the major
loop, then the avalanches occur in precisely the same time-order and it
is plausible to infer that the microscopic spatial evolution 
of the domains is also the same.
However, very recently it became possible to directly probe 
the precise spatial distribution of the magnetic domains 
using coherent x-ray speckle metrology (CXSM) instead of indirectly
inferring the domain configurations from the time
sequence of the avalanches \cite{3}.

Here we have applied this powerful new CXSM technique
to investigate the effects of
disorder on the domain evolution in a series of Co/Pt
multilayer samples with perpendicular magnetic anisotropy. In addition to
their possibly profound implications
for the existing theories about the
disorder and avalanches in 2d systems \cite{2}, such perpendicular magnetic films
are of intense current interest
because they promise significant increases in
magnetic storage density \cite{1}.

We address two longstanding questions in this letter:
(1) The existence of quasistatic major hysteresis loops
naturally raises the first question: How
is the microscopic magnetic domain configuration at one
point on the major loop related
to the configurations at the same point on the
major loop during subsequent
cycles?  We will call this effect major loop microscopic return point memory (RPM).
There are two limiting possibilities for this evolution:
(a)
The domains evolve in precisely the same
deterministic
way
during each
cycle; this is the prediction of all of the best
current (zero-temperature)
microscopic
theories \cite{2}.
(b) The domains evolve completely
differently during
each cycle;
their evolution is completely
non-deterministic;
this is the expectation for thermally
dominated systems.  There has been
surprisingly little theoretical work on
nonzero-temperature models.
We show below
that our samples exhibit behavior
intermediate
between these two
limits---their memories improve
with
increasing disorder, yet never become
perfect.
(2) The inversion symmetry
of the quasistatic major hysteresis
loops
through the origin
naturally
raises the second question: How are the
magnetic domains at one
point on the
major loop related to the domains at
the
complementary point 
during the same and during subsequent
cycles?
We will call this effect
major loop complementary point
memory
(CPM) \cite{5}.

\begin{figure}

\epsfxsize=7cm
\epsffile{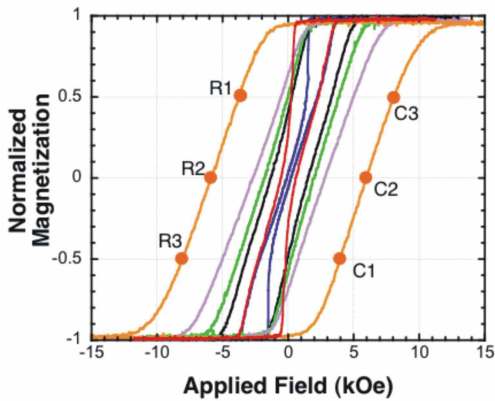}

\caption{(color online) Measured magnetic hysteresis
loops.
Starting at the origin,
and moving to increasing applied field $|H|$
the loops shown
are for samples 3, 7, 8.5, 10, 12, and 20.
The inversion symmetry through
the origin is shown.
}

\end{figure}

\begin{table}

\caption{\label{tab:table1}Co/Pt Sample
Characteristics
}

\begin{ruledtabular}

\begin{tabular}{cccc}
Sample\footnote{Growth pressure, milliTorr}
& $\sigma_{rms}$ \footnote{RMS Interfacial Roughness, nm}
& $M_{s}$ \footnote{Saturation Magnetization of Co, $ \rm{10^{3}emu/cm^{3}}$}
& $H_{c}$ \footnote{Coercive field, kOe}
\\
\hline
3  & 0.48 & $1.36 {\pm0.06}$ &  $0.08 {\pm0.01}$ \\
8.5  & 0.62 & $1.14{\pm0.05}$  & $1.3 {\pm0.2}$ \\
10  & 0.69 & $1.07{\pm0.05}$ & $1.6 {\pm0.2}$\\
12  & 0.90 & $1.10{\pm0.05}$ & $2.5 {\pm0.3}$\\
20  & 1.44 & $0.92{\pm0.04}$ & $5.9 {\pm0.2}$\\

\end{tabular}

\end{ruledtabular}

\end{table}

Our
samples were grown
by magnetron sputtering on
silicon-nitride-coated,
smooth, low-stress 1 cm
by 1
cm silicon wafers.  Each wafer had a 2 mm by 2 mm square 160-nm-thick
freestanding $\rm{SiN_x}$ membrane at its center. We used this thin
$\rm{SiN_x}$ window for our transmission
CXMS measurements.  Our samples all had
$\rm{ [Co (0.4~nm)/Pt (0.7~nm)]_{50} }$
with Pt buffer layers (20 nm) and Pt caps (2.3 nm).
Our sample numbers correspond to the pressure (in milliTorr)
of the argon gas that we used to sputter them. 
For low argon pressures, the
sputtered metal atoms arrive at the
substrate with considerable kinetic
energy which locally heats
and
anneals the growing film, leading to smooth
Co/Pt
interfaces \cite{6}.
For higher argon sputtering pressures, the
sputtered
atoms arrive
at the growth substrate with minimal kinetic energy,
resulting
in
rougher Co/Pt interfaces \cite{6}.

We measured the major hysteresis loops
for all of our samples using both Kerr and SQUID magnetometry.
The properties of the major loops show clear changes that are
related to the increasing roughness. Figure 1 shows that the
shape of the
major
loops change from being nucleation dominated to being
disorder dominated. 
Table 1 shows that the saturation magnetization
decreases as the microscopic disorder increases.
Our smoother samples
(3 and 7) exhibit soft
loops with low remanence
and distinct domain nucleation.
Between 7
and 8.5, there is
a transition to loops which do not show
a distinct domain nucleation region.  
Between 8.5  and 20, the ascending and
descending slopes of each loop remain the
same, but the loops gradually widen until
the full magnetic moment is left at remanence.
We show below that this change in character of the
loops coincides with the change in
their measured RPM and CPM.

In addition, we
also found via magnetometry
that all of our films
exhibit
{\it{perfect macroscopic major and minor loop}}
RPM and CPM \cite{3}.
This shows
that the ensemble-average magnetization is the same for
any given point on
a loop, but it does not prove that the individual
magnetic domains are
precisely the same!
We show below, in fact, that even for
our hard
magnetic
films with essentially
complete remanence,
that less than $ \sim 60\%$
of the domains are identical after reversal.

Our
experiments were performed at the Advanced Light
Source
at Lawrence
Berkeley
National Laboratory. We used linearly polarized
x-rays from the
third
and higher harmonics of the beamline 9
undulator.
The photon energy
was set to the cobalt $\rm{L_3}$ resonance at
$ \sim 778$ eV.
To
achieve
transverse coherence, the raw undulator beam was passed
through
a
35-micron-diameter pinhole
before being scattered in transmission
by the
sample.  The resonant
magnetic scattering was
collected by a soft
x-ray CCD
camera.  The magnetization of the film was varied by an
electromagnet,
which
applied fields perpendicular to the film.

The
effects of our controlled
disorder on the magnetic domains in real
space
and reciprocal
space are shown in Fig. 2.
The MFM images on the top
show that the domains form
characteristic worm-like labyrinths at low
disorder
but that these labyrinths become
increasingly indistinct as the structural
disorder is increased.
The magnetically scattered intensities
versus
the radial wavevector transfer are shown on the bottom.
They show
(1)
that the magnetic correlation length
decreases with increasing
disorder,
(2)that the magnetic domain size distribution
is well defined even though it is not
visible in the MFM images,
and 
(3) 
that increasing the disorder decreases the
characteristic domain size.

\begin{figure}

\epsfxsize=7.2cm
\epsffile{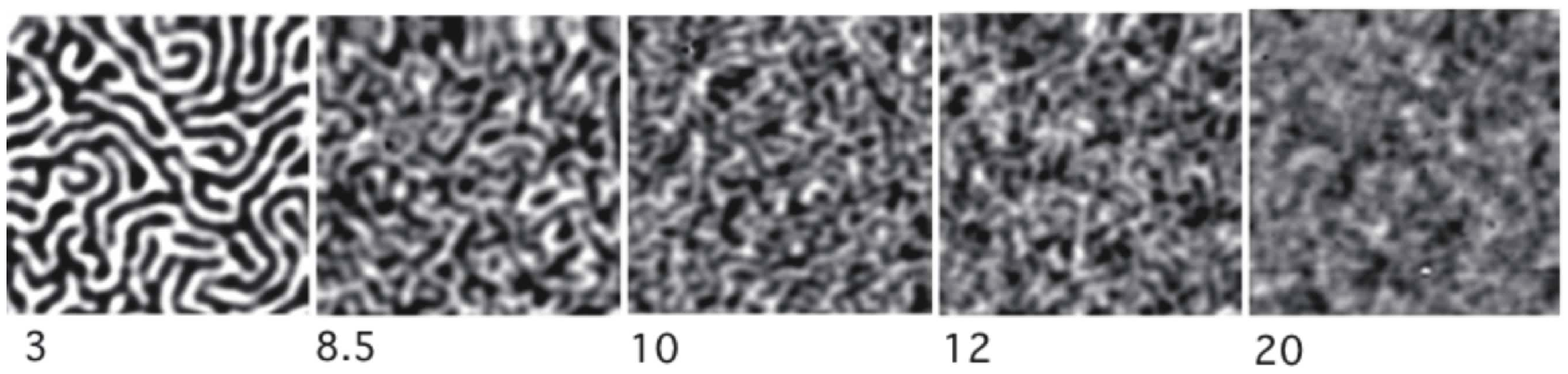}
\epsfxsize=6.8cm
\epsffile{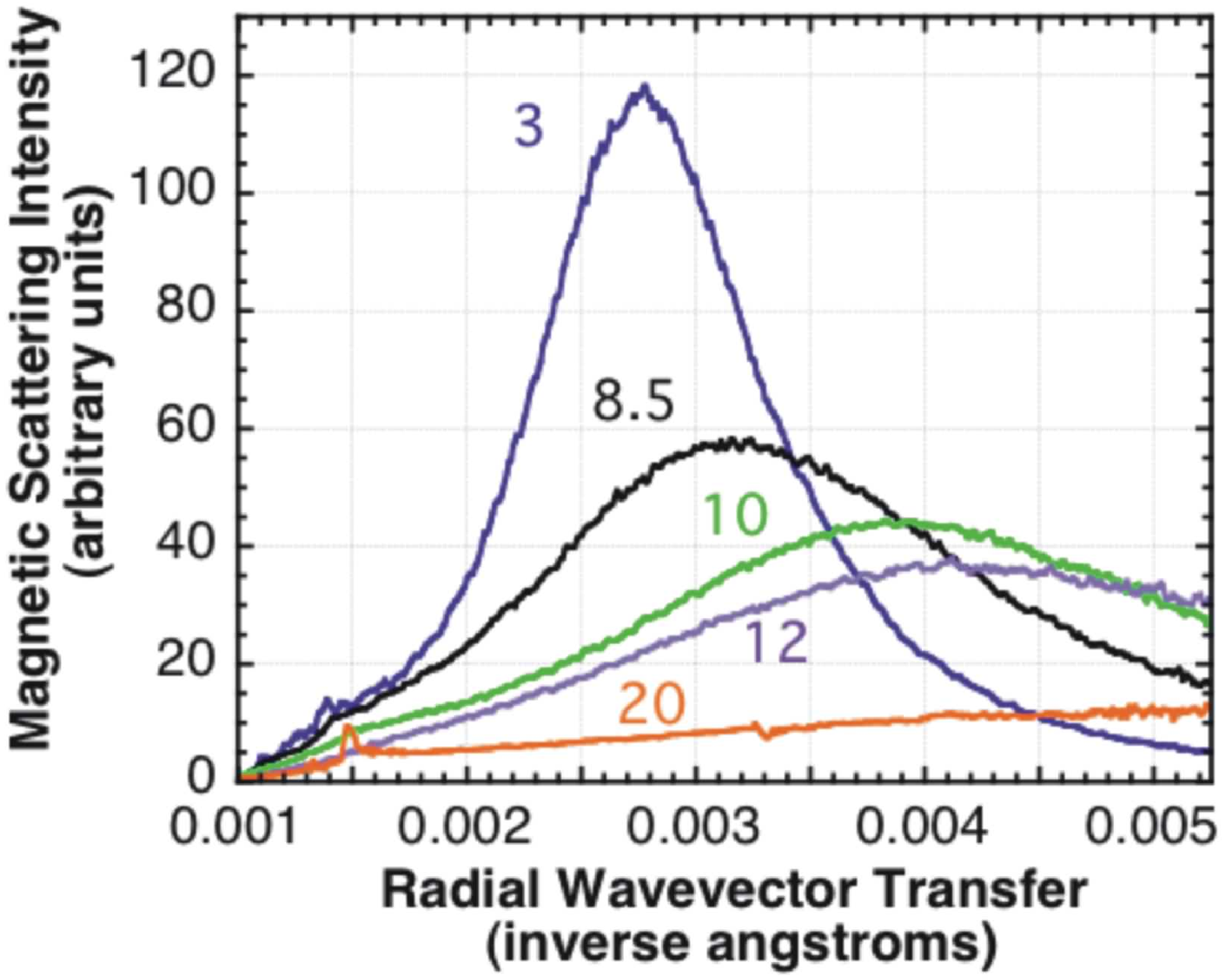}

\caption{(color online) Measured effects of the
argon sputtering pressure on the magnetic domains.
Top: Real space. Three square micron MFM images of the 
magnetic domain structure at remanence are shown.  
Note the apparent disappearance of the labyrinthine structure
as the disorder grows.
Bottom: Reciprocal space.  The radial profiles measured by
coherent x-ray magnetic scattering at the coercive point.
The labels (3, 8.5, 10, 12, and 20) indicate the argon sputtering pressure
in milliTorr during the growth of that film (see Table 1).
}
\end{figure}

We have previously shown that changes in the
magnetic domain structure can be measured via their magnetic speckle patterns \cite{3}.
To quantify the correlation between two speckle
patterns, we use image auto- and cross-correlation to extract the
generalized correlation coefficient $\rho$, which is unity
when the speckle patterns are identical and is zero
when the speckle patterns are completely uncorrelated \cite{3}.

The first question that we addressed was whether our samples exhibited
major loop microscopic RPM. To do so, we compared
two speckle patterns collected at the same point on
the major loop, but separated by one or more full excursions around
the major loop. For our smooth samples with soft loops,
we found  little correlation between these speckle patterns.
In sharp contrast, 
we always found strong correlations peaking near $\rho \sim 0.6$,
for our rougher, hard-loop samples.

The second question that we addressed was if
our samples exhibited major loop microscopic CPM.
To do so, we compared the speckle pattern collected
at one point on the major loop with speckle
patterns collected at the inversion-symmetric 
complementary point on subsequent cycles.  
Again, for our lowest disorder samples, we found no
correlation between these speckle patterns---zero CPM values.  
However, for the more disordered samples, we
found significant, non-zero CPM values that were consistently
smaller than the corresponding RPM values.

Our measured RPM and CPM values 
for sample 8.5 are shown in Fig. 3.  The data for
three sequential complete excursions
around the major loop are shown. For each value of the applied
field H, Fig. 3 shows the measured RPM and CPM values for that field.
There are several interesting trends:
(1) 
Neither the RPM value nor the CPM value depends on the number
of intermediate loops or on the choice of the starting half loop. 
This indicates that the deterministic components of the
RPM and CPM are essentially stationary.
This implies that the memory in our system is largely
reset by bringing the sample to saturation.
It also strongly suggests that the same disorder is
producing both the RPM and the CPM.
(2) 
Our RPM values are consistently larger than our CPM values.
(3) Both the RPM and the CPM values are largest near
the initial domain nucleation region, which occurs at about
-1 kOe for this sample.
This suggests that the subsequent decorrelation is produced by the
domain growth; this is confirmed by our magnetic x-ray imaging studies.
(4) 
Both the RPM and the CPM values decrease monotonically to their
minimum values near complete reversal which occurs at about 4 kOe 
for this sample. This again suggests that the decorrelation is
produced by the domain growth.

\begin{figure}
\epsfxsize=7.2cm
\epsffile{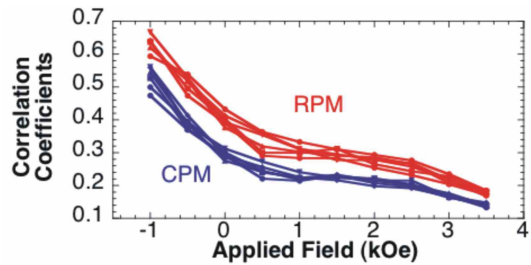}

\caption{
(color
online) Measured RPM and CPM
values versus the applied field for the 8.5 milliTorr sample.
}

\end{figure}

Figure 4 shows
our major loop microscopic RPM and CPM correlation coefficients,
measured at the coercive point for each sample, plotted versus our measured rms
roughness.  We measured the interfacial roughness for each sample,
by taking multiple AFM images and
calculating the rms variation of the measured heights.
We also verified that the roughness in our samples
was conformal and increased monotonically
with the argon sputtering pressure
by measuring their specular x-ray reflectivities using 0.154 nm x-rays.
As noted above, our two smooth samples have
essentially zero RPM and CPM values.
In contrast, all of our rough samples exhibit
non-zero RPM and CPM values that increase as the roughness increases.
This increase starts precisely where the
major loops change from being nucleation-dominated
to being disorder-dominated.

Our results are the first direct measurements of the effects of controlled
microscopic disorder on microscopic return point memory and complementary
point memory for major loops.  How can we understand our experimental
results in the light of the current microscopic disorder theories,
and in particular the discrepancy between the RPM and CPM values?

A widespread approach in the literature is to approximate the spins
as Ising, to use simple spin-flip dynamics, which is controlled 
solely by the energy of each spin, and to consider zero-temperature.
It is easy to see then,
that for models where the energy is unchanged when all the spins (and the external field) are reversed, that the configurations that the system goes
through on the ascending and descending branches of the major hysteresis
loop are precisely mirror images of each other, so that (at $T=0$) the RPM and CPM
values are both precisely unity.
This is the standard prediction for the $T=0$
random anisotropy, bond, and coercivity Ising models (RAIM, RBIM,
RCIM). On the other hand, for the $T=0$
random field Ising model (RFIM), only perfect
RPM is predicted. Note that both of these predictions are contrary to our experimental results;
it is therefore clearly important to include thermal effects in our simulations.

Within the Ising approximation with simple spin-flip dynamics
at nonzero temperature, our experimental 
result (that the CPM value is comparable but always smaller than the RPM value) 
implies that the disorder has a small component that breaks spin
reversal symmetry. This can be modeled by combining (for example)
the RFIM and the RCIM. Then our simulations using a dipolar $\phi^4$ model \cite{7}
show that a random field with an amplitude of only $\sim 4 \%$ of the spin-spin coupling can
produce an effect similar to what we see experimentally.
Our simulated results using this approximation are shown in Fig. 4.
Physically, the necessary local random fields could come either
from frozen impurity magnetic moments, or from 
a very wide distribution of domain coercivities and incomplete saturation.

\begin{figure}
\epsfxsize=7.2cm
\epsffile{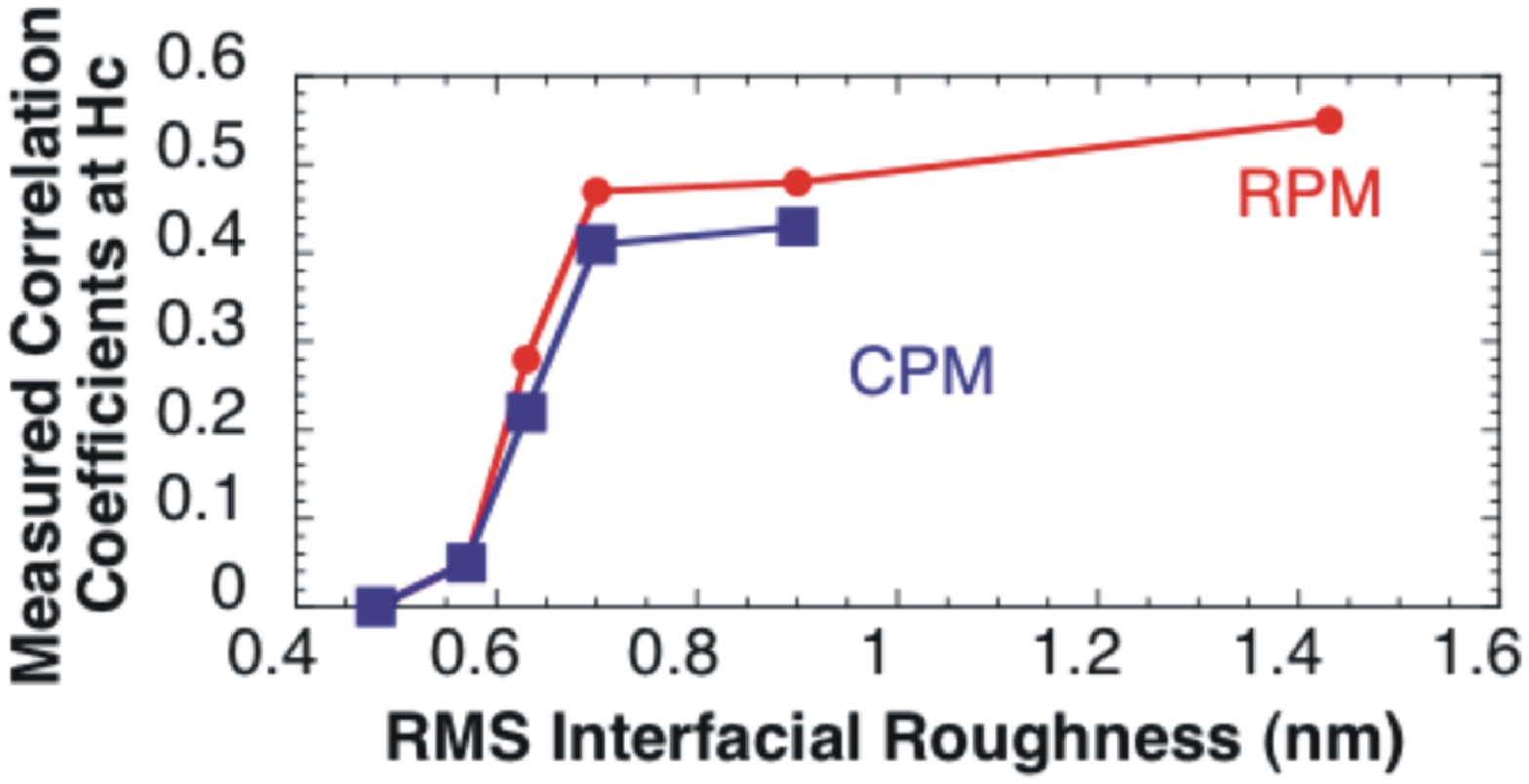}
\epsfxsize=7.2cm
\epsffile{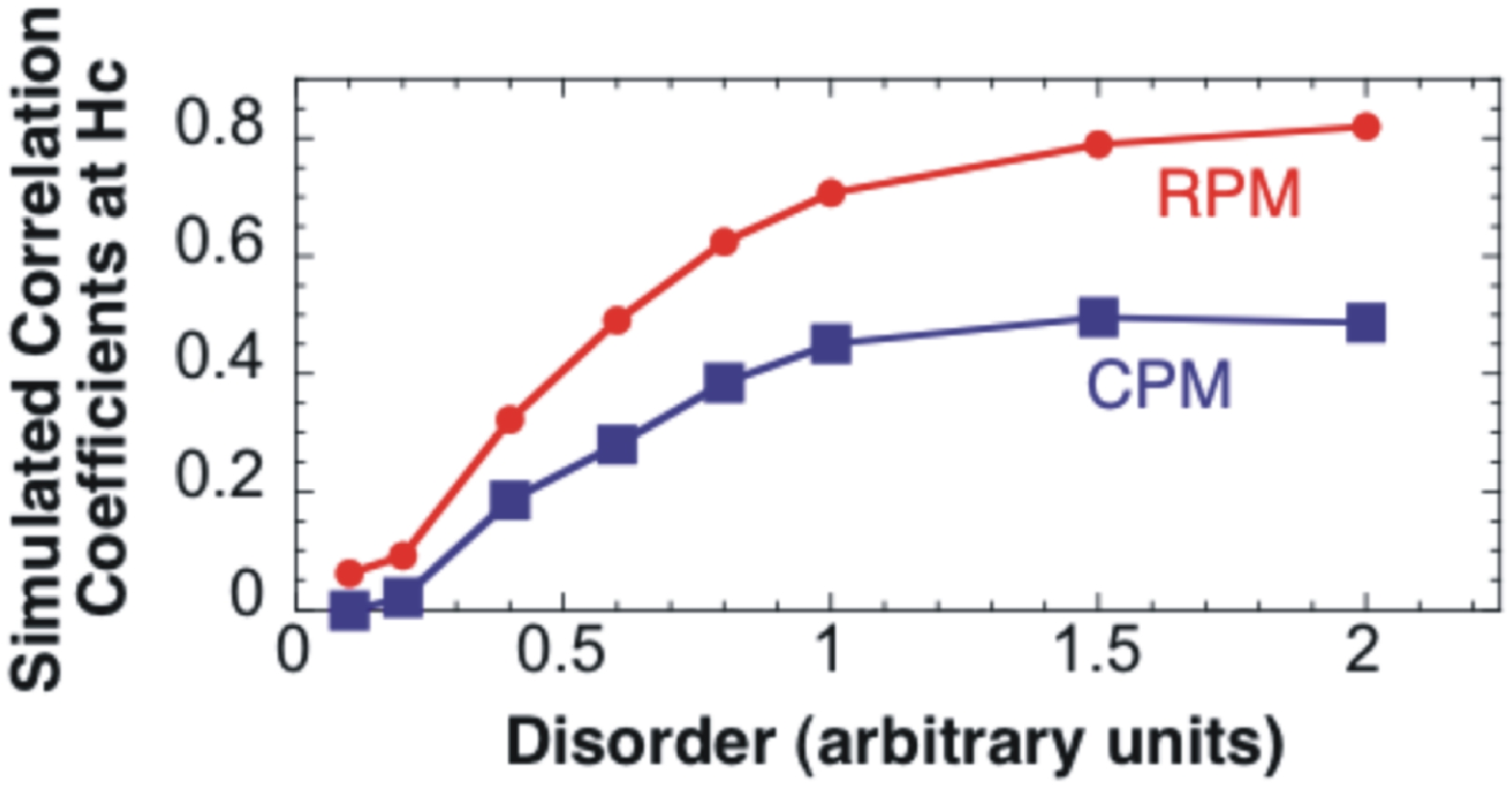}

\caption{
(color online)
RPM and CPM values at the coercive point versus disorder.
Top. Our measured RPM and CPM values versus the measured rms roughness. 
Bottom. The RPM and CPM values obtained from our nonzero-temperature numerical 
simulations that combine the RCIM and the RFIM. 
}

\end{figure}

However there is another potential origin for our observed RPM/CPM asymmetry. 
Even if the energy is spin-inversion symmetric, the dynamics
governing the spin reversal are not required to be. 
Our simulations of a spin-reversal-symmetric, disordered vector-spin model 
with strong Ising-like (locally varying) anisotropy, with the dynamics given 
by the standard Landau-Lifshitz-Gilbert-Bloch-Bloembergen formalism,
show dynamic symmetry breaking \cite{8}.
Note that then the time evolution of any spin is
given by a precessional term, which is unchanged under spin (and field)
reversal, and by a relaxational term, which is reversed.  Therefore, the overall
time evolution is not invariant under spin reversal, and the ascending
and descending microscopic states along the major hysteresis loop are not the same.

Our results have very strong implications. Within the widely used
approximation of Ising spins and simple spin-flip dynamics, they rule out all 
of the commonly used simple models---e.g., the RAIM, RBIM, RCIM and RFIM---and 
they necessitate a nonzero temperature combination of these. 
And if one goes beyond the simple Ising approximation,
we have found that it is possible for the
dynamics to break the spin-inversion symmetry even if the Hamiltonian is symmetric.

Our results show that there are still a variety of new interesting
experimental and theoretical questions in this classic mature subject:
What breaks the RPM/CPM symmetry?  Can the different RXIM models
be distinguished via the field dependence of their RPM and CPM values? 
It will be extremely interesting to see how the RPM and CPM values vary with the sample disorder and with sample temperature---particularly for technologically important magnetic 
memory materials.

We gratefully acknowledge support by the U.S.
DOE via DE-FG02-04ER46102,
DE-FG03-99ER45776, DE-AC03-76SF00098,
and DE-FG06-86ER45275 
and via LLNL-UCDRD.
We also gratefully acknowledge helpful discussions with
K.A. Dahmen and J.P. Sethna.
\dag OH was partially supported by the Deutsche Forschungsgemeinschaft
via HE 3286/1-1 and is currently at BESSY GmbH, Berlin, Germany.


\begin{references}





\bibitem{1}

A. Moser {\it et al.}, J. Phys. D {\bf 35}, R157 (2002); 
K. Ouchi, IEEE Trans. Magn. {\bf 37}, 1217  (2001).



\bibitem{2}
For RAIM see E. Vives and A. Planes, Phys. Rev. B {\bf 63}, 134431 (2001);
for RBIM see E. Vives and A. Planes, J. Magn. Magn. Mater. {\bf 221}, 164 (2000);
for RCIM see O. Hovorka and G. Friedman, cond-mat/0306300 (2003);
for RFIM see J.P. Sethna, K.A. Dahmen and O. Perkovic, cond-mat/0406320 (2004).



\bibitem{3}

M.S. Pierce {\it et al.}, Phys. Rev. Lett. {\bf 90}, 175502 (2003);
B. Hu {\it et al.}, Synch. Rad. News {\bf 14}, 11 (2001).





\bibitem{4}

S. Zapperi  and  G. Durin, Comp. Mat. Sci. {\bf 20}, 436 (2001);
J.R. Petta, M.B. Weissman and  G. Durin, Phys. Rev. E {\bf 57}, 6363 (1998);
D. Spasojevic {\it et al.}, Phys. Rev. E {\bf 54}, 2531 (1996).



\bibitem{5}

J.R. Hoinville {\it et al.}, IEEE Trans. Magn. {\bf 28}, 3398 (1992);
A. Jander {\it et al.}, IEEE Trans. Magn. {\bf 34}, 1657 (1998);
P. Fischer {\it et al.}, J. Phys. D {\bf 35}, 2391 (2002).



\bibitem{6}

E.E. Fullerton {\it et al.}, Phys. Rev. B {\bf 48}, 17232  (1993).



\bibitem{7}

E.A. Jagla, cond-mat/0402406 (2004).



\bibitem{8}

J.M. Deutsch, T. Mai, O. Narayan, cond-mat/0408158 (2004).

\end{references}
\end{document}